\documentclass [12pt,preprint]{aastex}
\usepackage{epsfig}
\begin{document}
\voffset-1cm
\newcommand{\gsim}{\hbox{\rlap{$^>$}$_\sim$}}
\newcommand{\lsim}{\hbox{\rlap{$^<$}$_\sim$}}

\title{Double-peak spectral energy density of GRBs\\
        and the true identity of GRB 031203}

\author{Shlomo Dado\altaffilmark{1} and Arnon Dar\altaffilmark{1}}

\altaffiltext{1}{dado@phep3.technion.ac.il, arnon@physics.technion.ac.il, 
dar@cern.ch.\\
Physics Department and Space Research Institute, Technion, Haifa 32000, 
Israel}

\begin{abstract}

A double-peak spectral-energy-density of $\gamma$-rays, similar to that
observed in blazars, is expected in $\gamma$-ray bursts (GRBs) produced in
supernova (SN) explosions. The second peak, which is formed by inverse
Compton scattering of the ambient SN light by cosmic ray electrons
accelerated by the jets from the SN explosion, has a much higher
peak-energy than the first ordinary peak. However, in X-ray flashes
(XRFs), which in the cannonball (CB) model are normal GRBs viewed farther
off axis, the first peak-energy is shifted to the soft X-ray region while
the second peak-energy moves to the MeV range. In far-off-axis GRBs, such
as GRBs 980425 and 031203, the second peak may have been confused with the
normal GRB peak. In most GRBs, which have been observed so far, the
$\gamma$ ray detectors ran out of statistics far below the second peak.
However, in bright GRBs, the two peaks may be resolved by simultaneous
measurements with SWIFT and GLAST.

\end{abstract}

\section{Introduction}

There is mounting evidence that long duration $\gamma$-ray bursts (GRBs)
and X-ray flashes (XRFs) are both produced by highly relativistic and
narrowly collimated jets ejected in supernova (SN) explosions akin to
SN1998bw (see, e.g. Dar 2004 and references therein).  Low-luminosity GRBs
and XRFs seem to be ordinary GRBs whose $\gamma$-rays are much softer and
their luminosity is much lower because they are viewed from angles
relative to the jet which are a few times larger than the viewing angles
of ordinary GRBs (see, e.g.  Dar \& De R\'ujula 2000,2003; Dado, Dar \& De
R\'ujula 2002,2004 and references therein). However, recently, the
European Space Agency satellite INTEGRAL discovered GRB 031203 (G\"otz et
al.~2004) at a redshift $z=0.1055$ (Prochaska et al.~2004), the second
nearest GRB after GRB 980425 at $z=0.0085$ (Galama et al.~1998), which
like GRB 980425 had extremely low lunminosity, but, contrary to the
expectation, its $\gamma$-rays were not much softer than those of ordinary
GRBs (Sazonov et al. 2004). GRB 031203 was produced in an SN very similar
to SN1998bw which produced GRB 980425 (e.g. Malesani et al.~2004 and
references therein). Consequently, the INTEGRAL observations were
interpreted as evidence of a different class of GRBs, which are
intrinsically very faint and include GRB 980425 and GRB 031203 (Sazonov et
al.~2004; Soderberg et al.~2004; Woosley et al.~2004) and are produced in
SNe akin to SN1998bw. 

The spectrum of the soft $\gamma$-rays emitted in GRB 031203, which was
measured with INTEGRAL, is well described by $dn_\gamma/dE\sim E^{-1.63\pm
0.06}$ between 20 and 400 $keV$ (Sazonov et al.~2004). But, its soft X-ray
fluence in the 0.2-10 $keV$ range, $F_{_X}\sim 2.6 \pm 1.3\, \times
10^{-7}\, erg\, s^{-1}\,,$ which was inferred (Watson et al.~2004) from
its X-ray dust-scattered echo measured by XMM-Newton (Vaughan et
al.~2004), is far above the INTEGRAL spectrum extrapolated to the soft
X-ray region.  It has been suggested that the soft X-ray fluence may have
been overestimated (Prochaska et al.~2004; Sazonov et al. ~2004).

However, in this letter we suggest an alternative interpretation of the
INTEGRAL and XMM-Newton observations, which is based on the cannonball
(CB) model of GRBs and XRFs (e.g. Dar \& De R\'ujula 2003; Dado et
al.~2004) and which removes the conflicts between the off-axis
interpretation of the unusually low luminosity of GRB 031203 and its high
peak-energy, and between its X-ray fluence extrapolated from the INTEGRAL
measurements (Sazonov et al.~2004) and that inferred from the XMM-Newton
measurements (Vaughan et al.~2004).  We also demonstrate that this
interpretation can explain other puzzling GRB observations.

The CB model assumes that in core-collapse SN, after the collapse of the 
stellar core into a neutron star or a black  hole, an accretion disk or
a torus is produced around the newly formed compact object, either by
stellar material originally close to the surface of the imploding core and
left behind by the explosion-generating outgoing shock, or by more distant
stellar matter falling back after its passage (De R\'ujula 1987). A
highly relativistic plasmoid (CB) of ordinary matter is emitted 
along the rotation axis when part of the
accretion disk falls abruptly onto the compact object, as observed in
microquasars (e.g.~Mirabel \& Rodrigez 1999; Rodriguez \& Mirabel 1999
and references therein). 
A CB contains a thermal plasma with a power-law tail of knocked-on
electrons from collisions with particles of the interstellar medium (ISM)
and swept-in ISM electrons, which are Fermi-accelerated and cool quickly
by synchrotron emission from the strong equipartition magnetic field in
the CB to a steady state distribution, $dn_e/dE\sim E^{-(p+1)}$ with $p\sim
2.2\,,$ in the CB rest frame. As the jet of CBs coasts through the ``ambient
light'' permeating the surroundings of the parent SN, the electrons
enclosed in the CBs Compton up-scatter photons to energies which, close to
the CBs direction of motion, correspond to the $\gamma$-rays of an
ordinary  GRB and less close to it, to the X-rays of an XRF
(Dar \& De R\'ujula 2003).

A CB also produces a narrow conical jet of high energy cosmic-ray
electrons (and nuclei) along its motion (e.g., Dar \& De R\'ujula 2003) .
These electrons are of two origins: swept-in ISM electrons which were
Fermi-accelerated in the CB and escaped out into the ISM and ISM electrons
which were scattered elastically by the CB. These cosmic-ray electrons
produce a second peak in the spectral energy flux of GRBs and XRFs at a
much higher energy, by inverse Compton scattering (ICS) of ``ambient
light'' permeating the surroundings of the parent SN.

Thus, the internal CB electrons and the external cosmic ray electrons
produce, by ICS of ambient light, a spectral-energy-density ($E^2\,
dn/dE$) of $\gamma$-rays with two peaks, like that observed in blazars
(e.g., Padovani \& Giommi 1995; Wehrle et al. 1998).  Normal GRBs have
their first peak-energy usually around a fraction of an MeV and then their
second peak-energy at a much higher energy. Gamma-ray detectors on board
satellites usually ran out of sensitivity/statistics well below the second
peak-energy. However, high-energy photons with a flux much larger than
that expected from the extrapolated decline of the first peak, have been
discovered in a few cases of very bright GRBs with instruments on board
CGRO (e.g. Hurley et al.~1994; Dingus et al.~1995; Gonzalez et al.~2003).
In this letter we demonstrate that such a high energy component is well
described by the CB model.  Moreover, in very low-luminosity GRBs/XRFs,
which are far off-axis GRBs, the first peak is in the $keV$ range while
the second peak is in the MeV range, i.e., the spacing between the peaks
is relatively small and both peaks, or a significant fraction of both, may
fall within the detection range of the ordinary XRF and GRB detectors. A
score of low-luminosity XRFs were detected by CGRO, BeppoSAX and HETE II.
But, in the case of HETE II, its FREGATE detector could go only to 400
keV.  BATSE on board CGRO and GRBM on board BeppoSAX usually ran out of
statistics/sensitivity well below the second peak, although a significant
emission in excess of that expected from a simple power-law extrapolation
of the low energy spectrum to higher energies was detected in some XRFs
(see e.g., Frontera et al. 2004a,b).  In this letter we first demonstrate
that the peak-energy and the isotropic radiation energy emitted in the
first peak satisfy well the simple correlation $(1+z)\,E_p\propto
[E^{iso}_\gamma]^{1/3}\, ,$ predicted by the CB model.  GRBs 980425 and
031203, which appear to be exceptions, could have been far-off-axis GRBs
with a relatively small spacing between their keV and MeV peaks, and
therefore appeared as GRBs with a ```normal'' peak-energy.  We predict
that simultaneous measurements with SWIFT and GLAST of the spectral energy
flux of ordinary GRBs will resolve it into two peaks which shift to
smaller energies during the GRB pulse.

\section{The first peak}

Let $\gamma$ be the Lorentz factor of a CB and 
$\delta=1/\gamma\, (1-\beta\, \cos\theta)$ be its Doppler factor when 
viewed from an angle $\theta$ relative to its motion.
In the CB model (Dar \& De R\'ujula 2003), 
the observed peak-energy of 
$\gamma$-rays produced at a redshift $z$ by ICS of thin thermal 
bremsstrahlung 
light around an SN, $dn_\gamma/dE \sim exp(-E/T)/E $
with a typical energy $\epsilon_\gamma\sim T\sim 1\,  eV\, ,$ 
is given by, 
\begin{equation}
E_p\approx  {\gamma\,\delta\over 1+z}\,T\, .
\label{Epeak1}
\end{equation}
Under the assumption of isotropic emission in the CB rest frame, Doppler 
boosting and relativistic beaming 
yield a  $\gamma$-ray fluence $F_\gamma$ of a GRB pulse, which is 
proportional to $\delta^3\, ,$  
\begin{equation}
F_\gamma\approx {\delta^3\, (1+z)\, E'_\gamma\over 4\, \pi\, D_L^2}\,,
\label{Fluence}
\end{equation}
where $E'_\gamma$ is the total $\gamma$-ray energy emitted 
in the CB rest frame.
Consequently,  $E^{iso}_\gamma$, the inferred `GRB isotropic 
$\gamma$-ray energy' in a GRB pulse, {\bf under the assumption
of isotropic emission in the observer frame}, is
\begin{equation}
E^{iso}_\gamma= {4\, \pi\, D_L^2\over 1+z}\, F_\gamma
 \approx \delta^3\, E'_{\gamma}\, .
\label{Eiso} 
\end{equation}

In the CB model, the predicted GRB spectrum from
ICS of ambient light with a thin thermal bremsstrahlung spectrum,
by the electrons inside the CB, is,
\begin{eqnarray} 
{ dN_\gamma[1]\over dE}
&\propto& \left({T_{eff}\over E}\right )^\alpha\; e^{-E/T_{eff}}+b\; 
(1-e^{-E/T_{eff}})\;
{\left(T_{eff}\over E\right)}^\beta\nonumber\\ \alpha&\approx&1\; ;
\;\;\;\;\;\;\;\; \beta={p+2\over 2}\approx 2.1\, , 
\label{totdist}
\end{eqnarray} 
where $T_{eff}=\gamma\, \delta\, T/(1+z)$ and $b$ is a
dimensionless constant. The values of
$\alpha$ and $\beta$ may deviate from their indicated  values, because the 
ambient radiation may deviate from a thin thermal bremsstrahlung, 
and the power-law index of
the accelerated and knocked-on electrons after cooling may be 
larger than 
$p+1=3.2$ and increase with time. Also 
the temperature of the ambient 
light seen by the CB is decreasing with distance, approximately as
\begin{equation}
T_{eff}(t)\sim T_{eff}(0)\,\{1-exp[-(t_0/t)^2]\},
\label{Tevol}
\end{equation}
where $t_0$ is a constant.
As was shown in Dar \& De R\'ujula 2003 and in Dado et al.~2004,
Eq.~(\ref{totdist}) is practically indistinguishable from the 
phenomenological Band
function (Band et al.~1993) and it is in good agreement with the measured
spectrum of the photons in the first peak of the 
spectral-energy-density of ordinary GRBs and XRFs. 

\begin{figure}[t]  
%\vskip -2cm
\begin{tabular}{cc} 
\hspace {1.7cm}
\epsfig{file=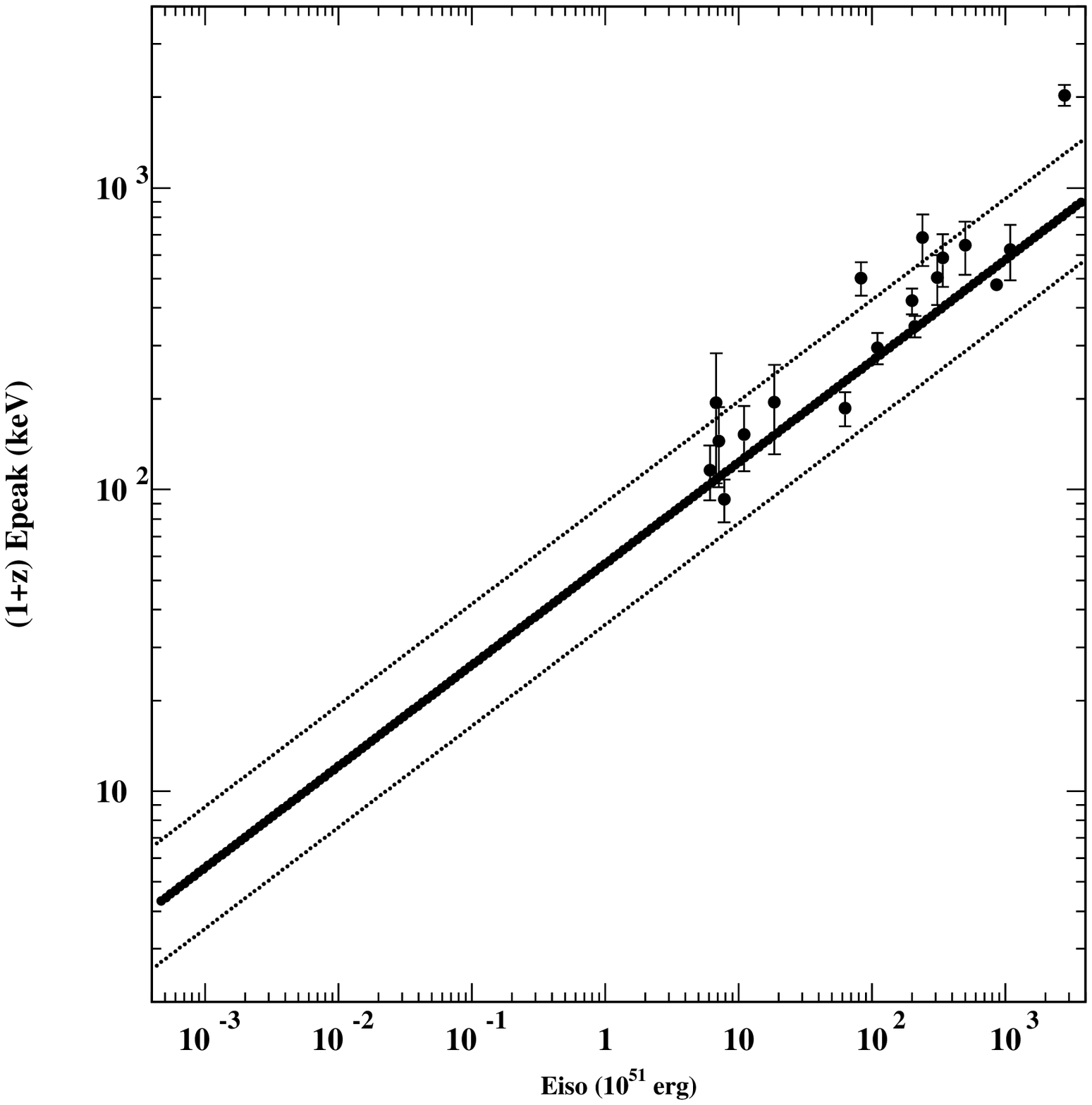, width=10cm}
%\vspace {-4cm}
\\
\hspace {1.7cm}
\epsfig{file=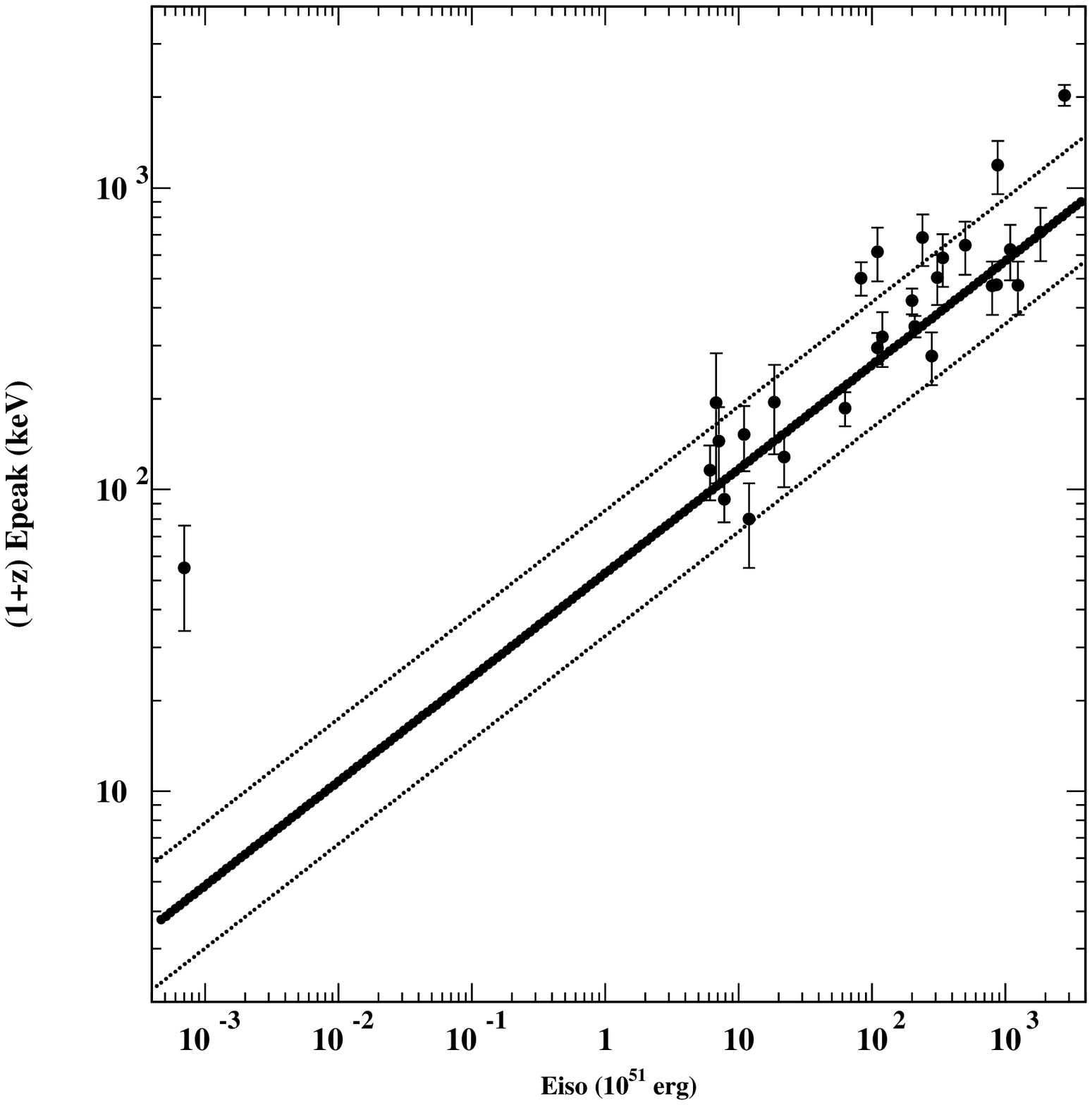, width=10cm}
%\vspace {-5cm}
\end{tabular}
\caption{The observed peak-energy corrected for redshift, 
$(1+z)\, E_p\, $ as function of the inferred isotropic radiation energy
emitted at redshift $z$ for GRBs/XRFs of known  redshift as compiled by
(a) Amati (2004) and by (b) Ghiralanda et al.~(2004). 
The thick lines 
are the best fitted power-laws: $A\, E^{0.334\pm 0.06}$ 
and $A\, E^{0.345\pm 0.07}\, ,$ respectively. The dotted lines 
border the 
estimated spread (a factor of $\sim 4$) in the isotropic radiation energy 
due to the spread in the number of CBs and their physical properties 
in GRBs. The point due to GRB 980425 (as inferred by Yamazaki et al. 2003) 
is far above the theoretical expectation.} 
\label{fig1ab}
\end{figure}

If core collapse SNe and their environments were all identical, and
if their ejected CBs were also universal in number, mass,
Lorentz factor and velocity of expansion, all differences
between GRBs would depend only on the observer's
position, determined by $z$ and the angle
of observation, $\theta$. For a distribution of Lorentz factors that
is narrowly peaked around $\gamma\simeq 10^3$,
the $\theta$-dependence is in practice the dependence on $\delta$,
the Doppler factor. Hence  Eqs.~(\ref{Epeak1}),(\ref{Eiso}) yield 
the correlation (Dar and De R\'ujula 2000; 2003),
\begin{equation}
(1+z)\,E_p\propto [E^{iso}_\gamma]^{1/3}\, .
 \label{EisoEp} 
\end{equation}
In Fig.~\ref{fig1ab} we plot $(1+z)\, E_p$ as function 
of $E^{iso}_\gamma$ for GRBs/XRFs of known redshift,
as compiled by  Amati  (2004) and by Ghiralanda et al.~(2004),
and the best fitted power-law,
$(1+z)\,E_p = a\, [E^{iso}_\gamma]^{\alpha}\,.$ 
The best fitted power-law index obtained with
the CERN standard 
program MINUIT, $\alpha=0.335\pm 0.06$
and $\alpha=0.345\pm 0.07$, for the Amati (2004) and 
Ghiralanda et al.~(2004) comilations, respectively, is in 
agreement with  Eq. (\ref{EisoEp}) but not with the original relation, 
$(1+z)\, E_p \sim a\, [E^{iso}_\gamma]^{0.52\pm 0.06}$ 
proposed by  Amati et al. (2002), which, as far 
as we know, has no theoretical derivation. 
We also plot in  Fig.~\ref{fig1ab} parallel lines 
between which the observed values are expected 
to be spread because of the spread in the `standard candle' properties.
The correlation predicted by the CB 
model is well satisfied except for GRB 980425 and GRB 031203.

\section{The second peak}

In the CB model, the ISM in front of the CBs is ionized by the 
beamed radiation from the highly relativistic CBs.
The turbulent magnetic fields in the CBs 
accelerate the ionized ISM particles, which they gather on their path,
to an initial distribution, $dn_e/dE\sim E^{-p}$ with $p\sim 2.2 $
in the CB rest frame. In a steady state situation, the electrons, which 
are trapped in the CB by its 
internal magnetic fields and 
cool quickly by synchrotron emission,  reach a 
distribution, $dn_e/dE\sim E^{-(p+1)}$, while the electrons which escape 
the CB must have  the  hard ``injection spectrum'', $dn_e/dE\sim E^{-p}\, .$
Their cooling time 
in the ISM is much longer than that of the electrons which are  trapped 
magnetically
inside the CBs, because the ISM magnetic field is smaller by many orders of
magnitude than that inside the CBs. 

In the CB rest frame, the Lorentz boosted ambient light 
undergoes Compton scattering from these two distributions and 
produces the first peak with the ``Band function'' shape and a hard tail    
at higher energies with practically a 
``time-independent'' spectral index 1.6,
\begin{equation}
{dn_\gamma\over dE} \propto  E^{-(p+1)/2}\sim E^{-1.6}\, .
\label{hardtail}
\end{equation}

The ISM electrons which are scattered elastically by the
highly relativistic CBs in the direction of the 
observer have an approximate lab
energy, $E_e\sim \gamma\, \delta\, m_e\, $. 
Because of their very large Lorentz factor, $\Gamma_e\sim \gamma\, \delta\,,$
the ambient photons which they scatter have much higher energies 
than those of the photons scattered by ``cold'' electrons in the CBs, and  
they are narrowly beamed along the electrons'direction of  
motion. Hence,
in the Thomson regime they have approximately the thin thermal bremsstrahlung
distribution of the ambient light boosted by $\sim 4\, \gamma^2\, 
\delta^2/3: $
\begin{equation} 
{E^2\, dN_\gamma \over dE}\propto E\, e^{-E/T_{eff}}\, ,
\label{thinbrem} 
\end{equation} 
where $T_{eff}=(4/3)\, \gamma^2\, \delta^2\, T/(1+z)\, .$
For $E \ll T_{eff}\, ,$ the low energy side of the spectral energy
density of this ``second peak'' has the simple shape, $E^2\, dN_\gamma/ dE
\propto E\, .$

During the GRB phase, the radius of a CB increases linearly with the
distance $x$ from the explosion site while the density of the ambient
light decreases like $1/x^2\, .$ Because of the slow cooling rate in the
ISM, the accumulated number of high energy electrons in the narrow beam
produced by the highly relativistic CB is proportional to the ISM mass
swept up by the CB. Thus, for a constant density-profile, the magnitude of
the spectral energy density, which is proportional to the product of the
swept-up mass and the density of the ambient light, increases linearly
with distance, whereas it decreases like $1/x$ for a density-profile of a
stellar wind which blows constantly and produces
a density $n_e\propto 1/x^2\,.$
During the GRB the deceleration of the CBs is negligible and the observer
time is proportional to the distance. Thus, for a constant density-profile
the spectral energy density increases with time like $ t\,,$ whereas it
decreases like $1/t$ for a ``windy'' density-profile.

In the Thomson regime, the peak-energies are given approximately by
$E_p \approx  T_{eff}/(1+z)\, .$ Consequently, 
$E_p[1]$ and $E_p[2]\,,$ 
the peak energies 
of the first and second peak, respectively,
are related through 
\begin{equation}
E_p[2]\approx {(1+z)\, [E_p[1]]^2\over T}\, .
\label{Ep1Ep2}
\end{equation}
For ordinary GRBs with $E_p[1]\sim 200\, keV$ and $z\sim 1\,,$ 
Eq.~(\ref{Ep1Ep2}) yields 
$E_p[2]\sim 100\, GeV\, ,$ whereas for a
very dim XRF with $E_p[1]\sim 1\, keV$ and $z<<1\, ,$ it yields, 
$E_p[2]\sim 1\, MeV\, .$ During each GRB pulse, both
peak-energies, being proportional to $T(t)\,,$ decrease
during a GRB pulse
like, $ T_{eff}(0)\, [1-exp[-(t_0/t)^2]]\, .$ 

\section{Comparison with observations}

There are very few cases where GRBs were measured over a very broad energy
range and allow a comparison between the CB model predictions and the
observations. Here we shall limit the comparison to two
GRBs of particular interest. 

\subsection{GRB  941017}

\begin{figure}[t]  
%\vskip -2cm
\hspace {1.7cm}
\epsfig{file=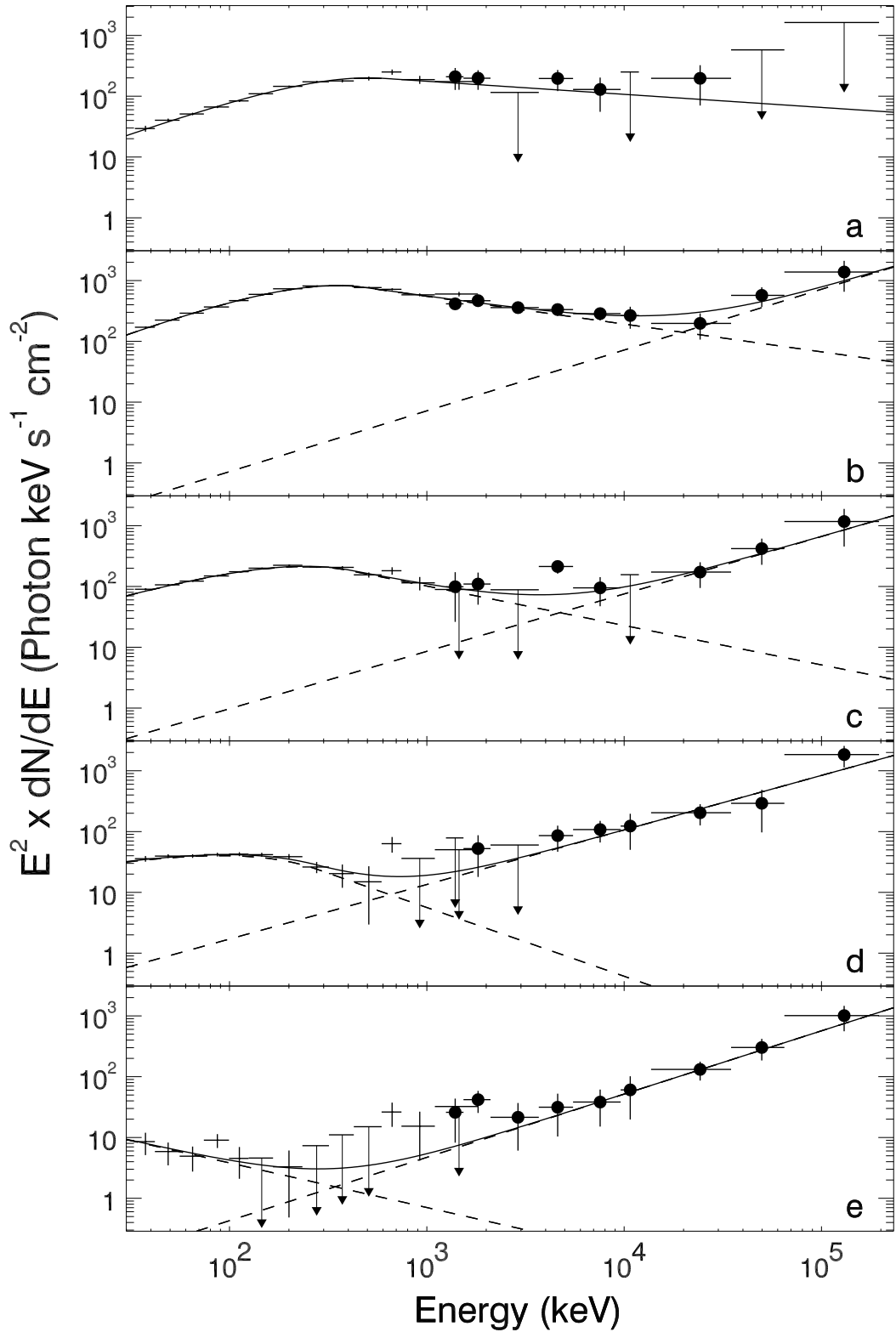 , width=15cm}
%\vspace {-5cm}
\caption{
The spectral energy density of GRB 941017 obtained by Gonzalez et al.~(2003) 
with  BATSE-LAD (crosses) and EGRET-TASC 
(filled circles) on board CGRO.  The CB model fit for the first 
peak is indistinguishable from the ``Band function'' fit (dashed line) by 
Gonzalez et al.~(2003). The rising side of the second peak is well
described by the CB model prediction as given by Eq.~(\ref{thinbrem}).
The fits are shown for the time intervals  (-18)-14s (a); 14-47s (b) ;  
47-80s (c);  80-113s (d); 113-211s (e).} 
\label{fig2}
\end{figure}

Gonzalez et al.~(2003) reported the discovery a high energy spectral
component in GRB 940117 with an energy flux density, $E^2\, dn/dE\sim E\,,$ 
peak energy  
$E_p[2]\geq 200\, MeV$ and a fluence $\geq$ 3 times that of the the first
normal GRB component. Their results are shown in Fig. \ref{fig2}  
borrowed from Gonzalez et al.~2003.
The theoretical lines are best fitted Band functions  
plus a  power-law contribution (Gonzalez et al.~2003). But, these 
lines are indistinguishable 
from the bes-fitted CB model double-peak  spectral-energy-density 
which is a sum of $E^2\, dn/dE$ with $dn/dE$
as given by Eq.~(\ref{totdist}) and of a second component which 
is described by  Eq.~(\ref{thinbrem}) with $E\ll T_{eff}[2]\,.$

\subsection{GRB 031203} 

\begin{figure}[t]  
%\vskip -2cm
\begin{tabular}{cc}
\hspace {1.7cm}
\epsfig{file=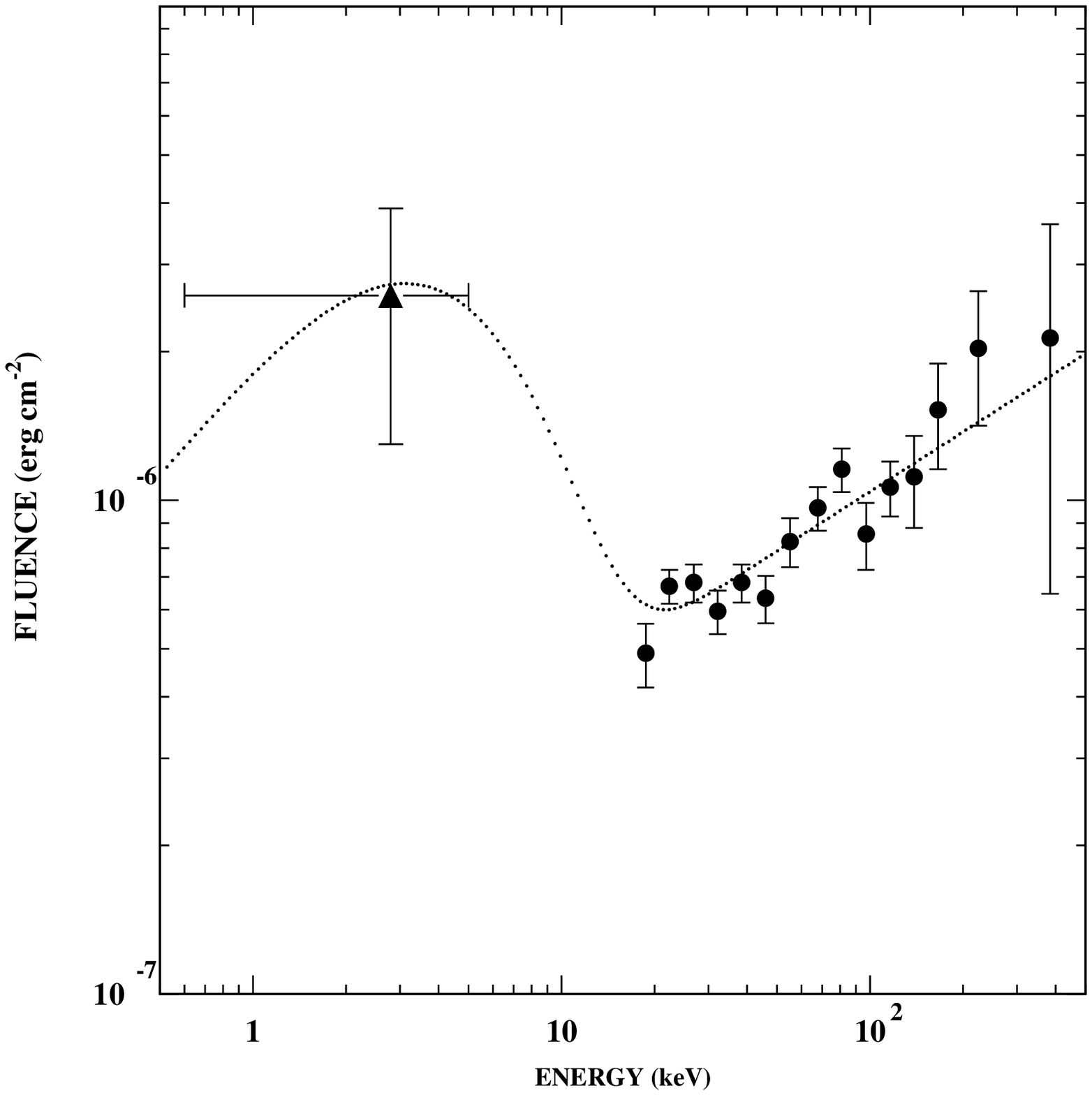, width=10cm}
%\vspace {-5cm}
\\
\hspace {1.7cm}
\epsfig{file=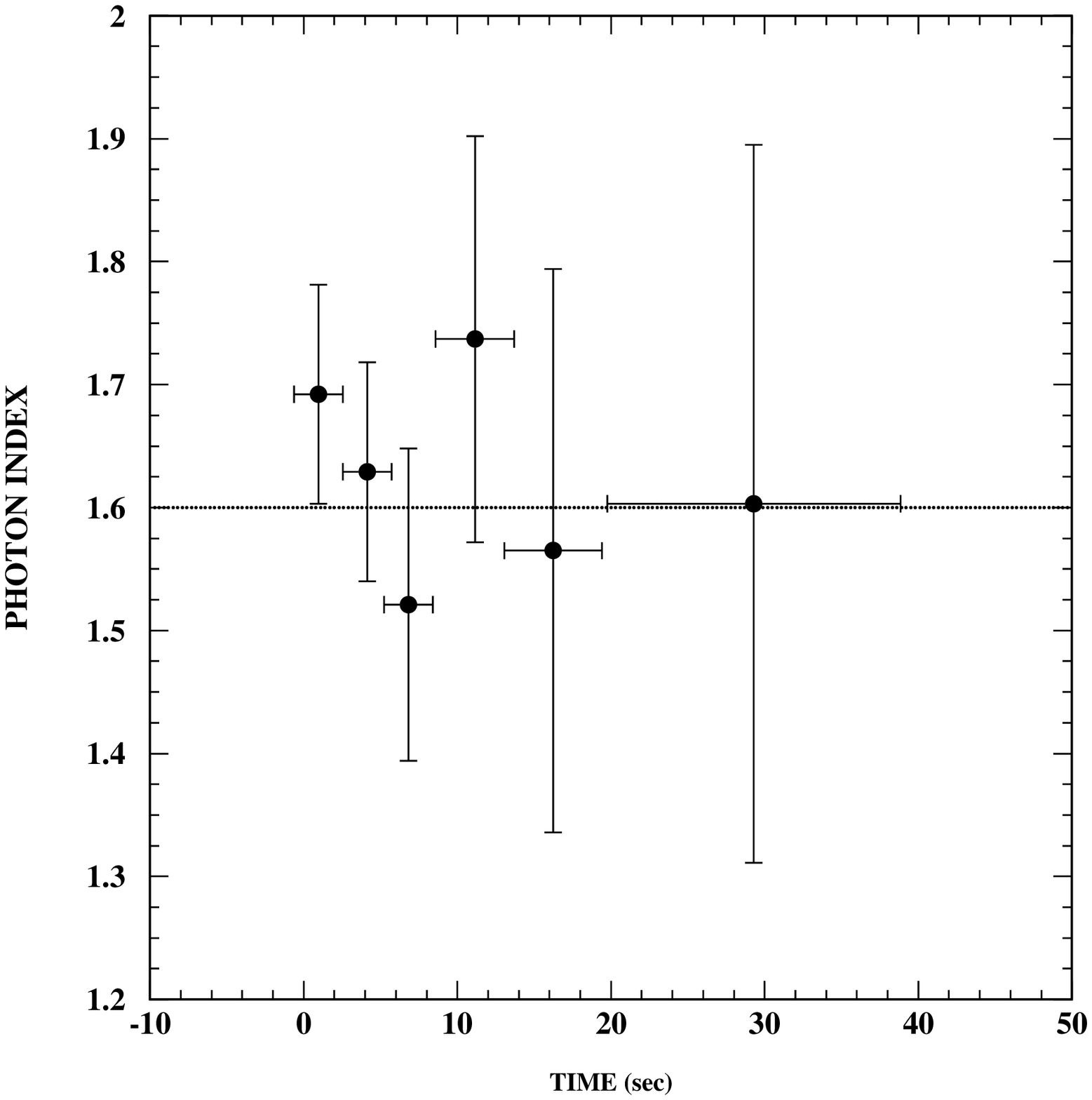, width=10cm}
\end{tabular}
\caption{Top:
Comparison between the observed spectral energy density of GRB 031203 
and a CB model fit given by Eq.~(\ref{totdist}) 
with $E_p\approx T_{eff}\approx 3\, keV$ and $b= 10^{-2}\, ,$
and a ``hard tail''
as given by  Eq.~(\ref{hardtail}). 
The X-ray fluence is that 
inferred by Vaughan et al.~(2004) from the dust echo observed with 
XMM-Newton. The INTEGRAL measurements are those reported by 
Sazonov et al.~(2004).
Bottom: Comparison between the the photon spectral index 
as obtained from 
single power-law fit to the photon spectrum 
measured with INTEGRAL (Sazonov et al.~2004)
and the CB model prediction for the hard-tail component 
in GRBs generated by cosmic ray electrons 
which were accelerated in the CBs.} 
\label{fig3}
\end{figure}

GRB 031203 (G\"otz et al.~2003) was produced in an SN explosion similar to
SN1998bw (Bersier et al.~2004; Cobb et al.~2004;  Malesani et al.~2004; 
Thomsen et al.~2004; Gal-Yam et al.~2004) and had an unusually low
inferred isotropic luminosity.  Its soft X-ray emission, which was
inferred from modeling its measured dust scattered echo by XMM-Newton,
yielded (Watson et al.~2004; Vaughan et al.~2004)), $F_{_X}\sim 2.6 \pm
1.3\, \times 10^{-6}\, erg\, cm^{-1}\,,$ in the 0.7-5 $keV$ range, far
above the extrapolation of the INTEGRAL spectrum. This can be seen from
the upper panel of Fig.~\ref{fig3} where we demonstrate a CB model 
fit to the spectral energy density of GRB 031203 which consists of 
two terms,
$E^2\, dn/dE$ with $dn/dE$ as given by Eq.~(\ref{totdist}) for a
low-energy peak with $E_p\approx T_{eff}\approx 3\, keV$ and $b= 10^{-2}\,
,$ and a hard-tail contribution as given by Eq.~(\ref{hardtail}). The
normalization of both components were adjusted to fit the observational
data. In the lower panel of Fig.~\ref{fig3}b we compare the photon
spectral index of the hard tail component, $(p+1)/2\approx 1.6\,,$ as
predicted by the CB model, Eq.~(\ref{hardtail}), and its measured value by
INTEGRAL (Sazonov et al.~2004). 

\section{Conclusions}

In the CB model, the internal population of electrons produce the ordinary
peak in the spectral-energy-density of GRBs/XRFs and the external
population of high energy (cosmic ray) electrons, which are accelerated by
the CBs, produce a second GRB peak at a much higher energy through inverse
Compton scattering of ambient light permeating the surroundings of the
parent SN.  We have shown that for GRBs with known redshift, usually the
peak-energy and the isotropic radiation energy of the first, ordinary peak
satisfy well the simple correlation $(1+z)\,E_p\propto
[E^{iso}_\gamma]^{1/3}\, ,$ predicted by the CB model. However, GRBs
980425 and 031203 have peak-energies much larger than those expected in
the CB model from their very low luminosities. But, in very low luminosity
GRBs and XRFs which, in the CB model, are ordinary GRBs viewed far off
axis, the second peak moves down to the MeV region and can dominate the
soft $\gamma$-ray region. We have shown that GRBs 031203 and GRB 980425
could have been such far-off-axis GRBs with their first peak in the soft
X-ray region but with a contribution from their second broad peak which
dominates their X-ray and $\gamma$-ray emission, as shown here explicitely
for GRB 031203. We have also shown that the high energy component which
was discovered in GRB 941017 is well explained by the second peak. 
Finally, we predict that simultaneous measurements with SWIFT and GLAST
will resolve the spectral energy flux of ordinary GRBs/XRFs into a double
peak spectrum whose peaks move to lower energies during the GRB pulse.

\noindent {\bf Acknowledgement:} 
Useful comments by Avishai Gal-Yam, Kevin Hurley and Simon Vaughan are
gratefully acknowledged. This researcch was supported in part by the Asher
Fund for Space Research at the Technion.

\end{document}